\documentclass[aps,prl,twocolumn,floatfix]{revtex4}
\usepackage{graphics,graphicx}
\usepackage{amsfonts}
\usepackage{bm,bbm}
\usepackage{mathrsfs}
\usepackage{calc}
\usepackage{epsfig}
\usepackage{float}
\usepackage{color}
\usepackage{epstopdf}

\def \be{\begin{equation}}
\def \ee{\end{equation}}
\def \ba{\begin{array}}
\def \ea{\end{array}}

\def\bpmat{\begin{pmatrix}}
\def\epmat{\end{pmatrix}}
\def\bmat{\begin{matrix}}
\def\emat{\end{matrix}}

\def\1{\mbox{1\hskip-.25em l}}

\def\beq{\begin{eqnarray}}
\def\eeq{\end{eqnarray}}
\def\be{\begin{equation}}
\def\ee{\end{equation}}

\def\beq{\begin{equation}}
\def\eeq{\end{equation}}

\begin{document}
\title{ Quantum correlation  effects on two successive  measurements  that are presented by non-commuting operators }
\author{Nimrod Moiseyev\footnote{nimrod@technion.ac.il} }
\affiliation{ Schulich Faculty of Chemistry, Solid State Institute, and Faculty of Physics, Technion-Israel Institute of Technology, Haifa 32000, Israel}
\date{\today}
\begin{abstract}
Two measurements  of $A$ and $B$  are carried out one after the other.  The measurements of $A$ are controlled by the parameter $\lambda_A$ in the Kraus operator, where the measurements of $B$ are controlled by the  parameter $\lambda_B$. Strong measurements  imply that the parameters in the Kraus operators approach infinite large values while weak measurements are carried out when the parameters approach zero. 
Here we prove that by repeating on the two successive measurements of $A$ and $B$  then:
(1)	Average over all measurements of $A$ is invariant of the measurement strength parameters $\lambda_A$ and $\lambda_B$. It implies that all surprising results obtained in weak measurements of $A$ are washed out when the average is taken.
(2)	If the operators $\hat A$ and $\hat B$ commute then the mean value of $B$ as obtained by taking the average of the results for $B$ over all measurements is invariant of $\lambda_A$ and $\lambda_B$. Moreover it is exactly equal to the expectation value of $\hat B$ as expected for strong measurements of $B$.
(3) If $\hat A$ and $\hat B$ do not commute \textit{and} another condition given in this paper is satisfied then the mean value of the results obtained for $B$ depends on the value of $\lambda_A$ and not on the value of $\lambda_B$.
An illustrative possible experiment to show the effect of the strength of the measurements of A on the results obtained for the measurements of B is given.
\end{abstract}
\pacs{????}
\maketitle

   Let us first explain the motivation of this work. The "unusual"  enhancement of a measured  quantity quantum weak measurements implies
that the value of a bounded operator $\hat A$ is measured not as the expectation
value in a pre-selected state $\hat A$ but also after post-selection by a state $\psi_{post}$ (see for example Ref.{\cite{Yakir6,Yakir8}}).  Weak measurements were already carried out in different physical contexts. See for example the experiments described in Refs.\cite{WM1}-\cite{WM14}.   When the result of the
measurement is registered by the shift of a pointer then this shift
can correspond to values which are far outside the spectrum obtained by projecting the pre-selected state on the eigenstates of $\hat A$. It was pointed out by Aharonov \textit{et-al}\cite{YakirJPB} that the large shifts of the pointer in weak measurements are associated with fast oscillations in an associate function involving the pre- and post-selected states. Berry and Shukra have shown that this happens if the initial pointer wavefunction is Gaussian (but not if it is exponential)\cite{BerryOSCILLATIONS}. This finding is in harmony with the  use of
Kraus operators as  pointers for measuring of A\cite{Kraus1},\cite{Kraus2}. Using the Kraus operator as a pointer it is quite straight forward to show that
the mean value of the results as registered by the pointer in weak measurement is exactly equal to the expectation value $\langle \psi_s|\hat A|\psi_s\rangle$. Namely, the mean value of the measured results are determined by the pre-selected state and there is no effect of the "unusual" results mentioned above on the the mean value as obtained by repeating the experiment  many times. The "unusual" results as obtained by specific post-selected states are washed out when the mean value of A is calculated. For quantum measurement and control see the book written by Wiseman and Milburn\cite{wiseman-BOOK}.

Our purpose here is to find out if there are situations where the fingerprints of these "unusual" results, that are obtained upon weak measurements, do exist also when the mean value of all results registered by the pointer is calculated.\\

 Here we wish to show that it is possible to design an experiment in a way that quantum correlations, due to the fact that two successive measurements are presented by non-commuting operators, can effect the mean value of a quantity that is taken after  weak measurements of another quantity were taken. We will show that quantum correlation effect might be observed only if the two operators that represent the measurement quantities do not commute. We will prove that the requirement of uncertain relations between the two measured quantities is a necessary but not a sufficient condition. In our studies presented here
 we follow previous studies where it has been assumed
that the system-detector interaction is instantaneous\cite{Mesnky}-\cite{Belzig}. This is to be distinguished from situations that are well described  by nonsymmetrized correlators that lie beyond the scope of
Markovian weak-measurement theory (see Ref.\cite{Belzig2} and references therein). Another point which should be clarified is the difference between our studies here of the weak-measurement  effect on  two successive measurements, which are presented by non-commuting operators, and previous studies of
measurements of non commuting operators in the context of quantum time correlation functions\cite{Belzig3},\cite{Belzig4}. There the measurements should be taken at different times. Our time independent studies presented below provide well defined conditions for observation of weak measurement effect when two successive measurements are taken for two quantities that are presented by non-commuting operators. \emph{As we will show the non-commuting of the operators is a necessary condition but not a sufficient one for the observation of a weak-measurement effect on the mean value of B when previously another quantity A has been detected. }

For the sake of clarity and coherent representation of our derivations we will represent again the system under study.
A quantum system is in a $|\psi_S\rangle$ which does not have to be an eigenstate of the Hamiltonian of the system under study. Yet it is a normalized function such that $\langle\psi_S|\psi_S\rangle =1$.
Two quantities are detected $A$ and $B$. Respectively they are associated with  non-commuting operators $\hat A$ and $\hat B$.
The quantum state $|\psi_S\rangle$ is not an eigenfuction of $\hat A$ or $\hat B$.
The spectra of these two operators are given by
\begin{eqnarray}
&& \hat A|a_n\rangle = a_n |a_n\rangle\,\,;\,\, n=1,2,... \nonumber \\
&& \hat B|a_m\rangle = b_m |b_m\rangle\,\,;\,\, m=1,2,...
\end{eqnarray}

Here box quantization is used for the continuous part of the spectrum of the two operators for the sake of simplification of the representation of our results and without loss of generality. Note that the spectrum of $\hat A$ and $\hat B$ can be degenerated as long as we keep all states to be orthonormalized.

We wish to detect the quantities A and B of the system under study when it is prepared to be in $|\psi_S\rangle$ state.
We have two available detectors: those which measure A only (the value of property A that is detected is denoted here as $a$ which does not have to be one of the eigenvalues of $\hat A$), and those which measure B only (the output of this detector is denoted here by $b$). We assign the values detected by these kind of detectors as $a,b$. We assume that any change in the state of the system under study is due to the operations of the detectors  of A and B only.
Every one of these detectors has a knob that determines the strength of the measurement. The device that measures A has a knob that controls the  measurement strength  parameter $\lambda_A$, and the device that measures B has a knob that controls the measurement strength parameter $\lambda_B$. In the limit of $\lambda_{A/B}\to \infty$ strong measurement is taken where the A detector yields only one of the eigenvalues of $\hat A$ (i.e., $a=a_n; \, n=1,2,...$); and detector B results in one of the eigenvalues of $\hat B$. The outputs of the two detectors are as follow:
detector A has the output $-\infty \le a \le +\infty$, and detector B has the output $-\infty  \le b \le  +\infty$.\\
 In the case the detector A measures the value $a$ the system under study collapses to a state which is given by  $$|\Psi_{A}(a)\rangle=\hat K_a^{\lambda_A}|\psi_S\rangle$$  where,  $$\hat K_a^{\lambda_A}=(\frac{2\lambda_A}{\pi})^{1/4}e^{-\lambda_A(a-\hat A)^2}$$. $a$ is the value detected by the first measurement where $\lambda_A$ ia the strength measurement parameter as mentioned above. In the case that detector B is applied on the system and the value for property B is measured  to be $b$  then the system after the detection collapses to a state which is given by, $$|\Psi_{B}(b)\rangle=\hat K_b^{\lambda_B}|\psi_S\rangle$$ where,  $$\hat K_b^{\lambda_B}=(\frac{2\lambda_B}{\pi})^{1/4}e^{-\lambda_B(b-\hat B)^2},$$ and  $\lambda_B$ is the measurement strength parameter of the second detector. $\hat K_a^{\lambda_A}$ and $\hat K_b^{\lambda_B}$  are the Kraus operators\cite{Kraus1},\cite{Kraus2} which provide the states of the system after taking a measurement of A and B.
  Here $a$ is the shift of the pointer of the detector that measure A and $b$ is the shift of the pointer of the detector that measures B. The question of the validity of the Kraus operators in the description of the "collapse" of the state as the detections are taken is out of the scope of this paper.

Let us first calculate the mean value of A (similarly for B) as obtained by the use of a detector with a strength measurement parameter $\lambda_A$. The mean value of $A^k$ (i.e., $k=1$ is the mean value of A and for the calculations of standard deviation of our measurement we need to calculate also the mean of $A^2$) as measured by the detector is given by,
\begin{eqnarray}
&& \bar{A^k}(\lambda_A)=\int_{-\infty}^{+\infty} a^k \langle\Psi_{A}(a)|\Psi_{A}(a)\rangle = \\ && \sum_n[(\frac{2\lambda_A}{\pi})^{1/2}\int_{-\infty}^{+\infty} a^k e^{-2\lambda_A(a-\hat a_n)^2}da]\times \nonumber \\ &&\langle \psi_S|a_n\rangle\langle a_n|\psi_S\rangle =\nonumber \\ && \sum_n[(\frac{2\lambda_A}{\pi})^{1/2}\int_{-\infty}^{+\infty} (\xi+a_n)^k e^{-2\lambda_A\xi^2}d\xi]\langle \psi_S|a_n\rangle\langle a_n|\psi_S\rangle \nonumber
\end{eqnarray}
For $k=1$ we get that that the mean value of A does not depend of the measurement strength parameter of the detector,
\begin{eqnarray}
&& \bar{A}(\lambda_A)= \sum_n a_n\langle \psi_S|a_n\rangle\langle a_n|\psi_S\rangle=\langle \psi_S|\hat A|\psi_S\rangle
\end{eqnarray}
which is the expectation value as defined in QM for strong measurements.\\ For $k=2$ we get that
\begin{eqnarray}
&& \bar{A^2}(\lambda_A)=\sum_n a_n^2\langle \psi_S|a_n\rangle\langle a_n|\psi_S\rangle + \\ && \sum_n[(\frac{2\lambda_A}{\pi})^{1/2}\int_{-\infty}^{+\infty} \xi^2 e^{-2\lambda_A\xi^2}d\xi]\langle \psi_S|a_n\rangle\langle a_n|\psi_S\rangle \nonumber
\end{eqnarray}
Since
\begin{eqnarray}
&&\int_{-\infty}^{+\infty} \xi^2 e^{-2\lambda_A\xi^2}d\xi=-\frac{1}{2}\frac{\partial}{\partial \xi}\int_{-\infty}^{+\infty} e^{-2\lambda_A\xi^2}d\xi=\nonumber \\ && -\frac{1}{2}\frac{\partial}{\partial \xi} (\frac{2\lambda_A}{\pi})^{-1/2}=\frac{1}{2\lambda_A}
\end{eqnarray}
we get that
\begin{eqnarray}
&& \bar{A^2}(\lambda_A)=\langle \psi_S|A^2|\psi_S\rangle + \frac{1}{2\lambda_A}
\end{eqnarray}

The standard deviation of A is defined as usual by $\sqrt{\bar{A^2}-(\bar{A})^2}$.
Therefore, the standard deviation of A as defined in quantum mechanics textbooks is given by
 \begin{eqnarray}
&& \Delta A(\lambda_A)=\sqrt{\Delta^2 A(\lambda_A\to\infty)+\frac{1}{2\lambda_A}} \\ && \simeq \Delta A(\lambda_A\to\infty)+\left(\frac{1}{4\Delta A(\lambda_A\to\infty)}\right)\frac{1}{\lambda_A}+...\nonumber
\end{eqnarray}

As one can see the standard deviation of A is increased as the measurement becomes weaker,  $$ \Delta A(\lambda_A\to 0)\simeq \sqrt{\frac{1}{2\lambda_A}}\to\infty,$$  although the mean value of the measurements does not dependent on the measurement strength parameter and gives the standard expectation value.

\textit{The question we address  here is whether  we can find situations where the  so called "unexpected/surprising" results are obtained even from the calculations of the mean value of measurements of one quantity.  Moreover, these "unuusal/surprising" results are obtained  \textbf{only} due to the uncertainty relations between the operators that represent these measurements.}

It is important to notice that in our studies the time of operation of every one of the detectors is assumed to be short as possible. Namely, $$\Delta t_{measurement} \to 0$$ where, $$\lambda_{A/B}\equiv \int_0^{\Delta t} \lambda_{A/B} (t))dt.$$ The measurement strength parameters $\lambda_A$ and $\lambda_B$ are held fixed when every one of the protocols mentioned below are taken.

\subsubsection {Sequential measurements of two quantities: first A and shortly afterwards B}
 Prior to the measurements the system is an state $|\psi_S\rangle$. Resulting of the first measurement of A the system "collapses" to $ |\psi_A^{collapse}(a,\lambda_A)\rangle= \hat K_a^{\lambda_A}|\psi_S\rangle$.  Shortly afterwards of this measurement we do another measurement of B. We assume that the second measurement is taken when the system is in $|\psi_A^{collapse}(a,\lambda_A)\rangle$ state. Therefore, after the second measurement of B the system "collapses" into a new state $|\psi_{AB}^{collapse}(a,b,\lambda_A,\lambda_B)\rangle=\hat K_b^{\lambda_B}|\psi_A^{collapse}(a,\lambda_A)\rangle$. In order to simplify the notation we will label this state below as $|\Psi_{AB}(a,b)\rangle$,
\begin{eqnarray}
&& |\Psi_{AB}(a,b)\rangle =\hat K_b^{\lambda_B} \hat K_a^{\lambda_A}|\psi_S\rangle \\ && =(\frac{2\lambda_A}{\pi})^{1/4}\hat K_b^{\lambda_B} e^{-\lambda_A(a-\hat A)^2}|\psi_S\rangle \nonumber  \\
 &&= (\frac{2\lambda_A}{\pi})^{1/4}\hat K_b^{\lambda_B}\sum_n e^{-\lambda_A(a-a_n)^2}|a_n\rangle\langle a_n|\psi_S\rangle \nonumber \\
 && =(\frac{4\lambda_A\lambda_B}{\pi^2})^{1/4}\times \nonumber \\ && \sum_{n,m} e^{-\lambda_A(a-a_n)^2-\lambda_B(b-b_m)^2}|b_m\rangle \langle b_m|a_n\rangle\langle a_n|\psi_S\rangle \nonumber
 \end{eqnarray}

where $a$ and $b$ are the shifts of the pointers of the two detectors that measure A and B.

 $P_{\lambda_A,\lambda_B}(a,b)$ is the probability to measure $a$ by the detector A and measure $b$ by the detector B when the measurement strength parameters are $\lambda_A$ and $\lambda_B$:

 \begin{eqnarray}
&&P_{\lambda_A,\lambda_B}(a,b)=\langle \Psi_{AB}(a,b)|\Psi_{AB}(a,b)\rangle= (\frac{4\lambda_A\lambda_B}{\pi^2})^{1/2} \times \nonumber \\ && \sum_{n,n'}\langle \psi_S|a_{n'}\rangle \langle a_n|\psi_S\rangle {e^{-\lambda_A[(a-a_n)^2+(a-a_{n'})^2]}} \times \nonumber \\ &&\sum_{m}e^{-\lambda_B(b-b_m)^2}\langle  b_m|a_n\rangle \langle a_{n'}|b_{m}\rangle
\end{eqnarray}

We might note in passing that in the limit of two strong measurements (first measuring A and then imminently measure B)

 \begin{eqnarray}
&& P_{\lambda_A\to\infty ,\lambda_B\to\infty}(a=a_{n_0},b=b_{m_0})= \nonumber \\ && |\langle a_{n_0}|\psi_S\rangle|^2 |\langle a_{n_0}|b_{m_0}\rangle|^2
\end{eqnarray}

which may be much larger than direct measurement of B. For example it may happen that $P_{\lambda_B\to\infty}(b=b_{m_0})=|\langle b_{m_0}|\psi_S\rangle|^2=0$
while $P_{\lambda_A\to\infty ,\lambda_B\to\infty}(a=a_{n_0},b=b_{m_0})\approx 1$ (smaller than unity). This situation implies that $|\langle \psi_S|a_{n_0}\rangle|^2\approx 1$ and also $|\langle a_{n_0}|b_{m_0}\rangle|^2 \approx 1$.
As an example we associate here the measurements of $a_{n_0}$ with the measurements of the positions of the quantum particles and $b_{m_0}$ with the measurements of the momenta. $|\langle x_{n_0}|\psi_S\rangle|^2\approx 1 $ and the $m_0$ component of the Fourier transform of $\psi_S(x)$ is out of the band limit of $\psi_S(x)$, i.e., $b_{m_0}=\hbar m_0$ is out of the Fourier transform limit of $\psi_S(x)$ and $\langle b_{m_0}|\psi_S\rangle = 0$.
The discrete  spectrum of the position operator is $a_n\equiv x_n=\Delta_x\times n$ where $n=0,\pm 1,\pm 2,...$  and  $\Delta_x<<1$. The corresponding eigenstates of the position operator are associated with a set of orthonormal functions that at $x=x_n$ get the value of unity,  $|x_n\rangle = 1$, and get the values of  $|x_n\rangle=sinc((x-x_n)/\Delta_x)$ elsewhere, ($sinc(\xi)=[\sin(\pi \xi)]/(\pi \xi)$). The Fourier transform  of a $sinc$ function is a rectangular with an one unity high and a very large width. Namely, the momentum obtained by taking the transform Fourier  is varied from $-\hbar\Delta_x/2 $ to $+\hbar\Delta_x/2$ where for example $\hbar\Delta_x/2 = b_{m_0}$. Consequently, $|\langle a_{n_0}|b_{m_0}\rangle|$ can be close to unity.  Our use of  the narrow window along the position direction presented by the $sinc$ functions to get a Fourier component that it is out of the band limit of $\psi_S(x)$ is very similar to the association of superoscillations with quantum weak measurements. Superoscillations implies that there are
regions where a band-limited function  varies
faster than the fastest of its Fourier components. See for example Ref.\cite{Berry-NM} and references therein.

Now we will study the effect of two successive weak measurements where $\lambda_A\to 0$ and also $\lambda_B\to 0$ on the probability to measure the values $a$ and $b$ by successive measurements of A and B,
 \begin{eqnarray}
&& P_{\lambda_A\to 0,\lambda_B\to 0}(a,b)\simeq \sqrt{\lambda_A\lambda_B}-\\ &&  \sum_{n,n',m}\langle \psi_S|a_{n'}\rangle\langle a_{n'}|b_m\rangle\langle b_m|a_n\rangle  \langle a_n|\psi_S\rangle \times \nonumber \\ && [\lambda_A^{3/2}\lambda_B^{1/2}[(a-a_n)^2+(a-a_{n'})^2]+  \lambda_B^{3/2}\lambda_A^{1/2}(b-b_m)^2]\nonumber
\end{eqnarray}

When $\lambda_A=\lambda_B\equiv\lambda$ in the limit of $\lambda\to 0 $ we get that the \textit{local maximal value of $P_{\lambda_A=\lambda_B\to 0}(a,b)$ for any value of "a" and "b"} is at
$$\lambda=\frac{1}{2C(a,b)}$$ where
\begin{eqnarray}
&&C(a,b)=\sum_{n,n',m}\langle \psi_S|a_{n'}\rangle\langle a_{n'}|b_m\rangle\langle b_m|a_n\rangle  \langle a_n|\psi_S\rangle \times \nonumber \\ &&[(a-a_n)^2+(a-a_{n'})^2+ (b-b_m)^2]>0
\end{eqnarray}

 This local maximum in the probability to measure any value of $a$ and $b$ ( $C>0$) at weak measurements of A and B  can be  large (in comparison with the probability to measure the values $a=a_{n_0}$ and $b=b_{m_0}$ in two successive  strong measurements of A and B) when C gets large positive values.
 For example let us consider the case where $P_{\lambda_A\to\infty ,\lambda_B\to\infty}(a=a_{n_0},b=b_{m_0})\simeq 0$.  Nevertheless,
 \begin{eqnarray}
&&C(a_{n_0},b_{m_0})=\sum_{n,n',m}\langle \psi_S|a_{n'}\rangle\langle a_{n'}|b_m\rangle\langle b_m|a_n\rangle  \langle a_n|\psi_S\rangle \times \nonumber \\ &&[(a_{n_0}-a_n)^2+(a_{n_0}-a_{n'})^2+ (b_{m_0}-b_m)^2]>0
\end{eqnarray}

 due to the weak measurements for which $\langle \psi_S|a_{n'=n}\rangle \langle a_n|\psi_S\rangle \simeq 1$ and $\langle a_{n'=n}|b_m\rangle\langle b_m|a_n\rangle \approx 1$. Again we use here the fact that when B is associated with the measurement of the momentum and A with the position the Fourier transform of a $sinc$ function has a very broad band.

 For the sake of complementary of our discussion on two successive measurements of A and B  below we will prove that regardless of the measurement strength parameters  $$\int_{-\infty}^{+\infty} da db \langle \Psi_{AB}(a,b)|\Psi_{AB}(a,b)\rangle =1.$$
For the sake of clarity we represent the algebraic manipulations (even though they are simple)
\begin{eqnarray}
&&\int_{-\infty}^{+\infty} dadb \langle \Psi_{AB}(a,b)|\Psi_{AB}(a,b)\rangle= \nonumber \\ &&(\frac{2\lambda_A}{\pi})^{1/2}\sum_{n,n'}[\int_{-\infty}^{+\infty}da e^{-\lambda_A[(a-a_n)^2+(a-a_{n'})^2]}]\times \nonumber \\ &&\sum_m\langle  b_m|a_n\rangle\langle a_n|\psi_S\rangle\langle \psi_S|a_{n'}\rangle \langle a_{n'}|b_m\rangle \nonumber \\ &&=(\frac{2\lambda_A}{\pi})^{1/2}\sum_{n,n'}[\int_{-\infty}^{+\infty}da e^{-2\lambda_A[(a-\frac{a_n+a_{n'}}{2})^2]}]\times \nonumber \\ && e^{-\lambda_A\frac{(a_n-a_{n'})^2}{2}}\sum_m\langle a_{n'}|b_m\rangle\langle  b_m|a_n\rangle\langle a_n|\psi_S\rangle\langle \psi_S|a_{n'}\rangle  \nonumber\\&& =\sum_{n,n'}e^{-\lambda_A\frac{(a_n-a_{n'})^2}{2}} \langle  a_{n'}|a_n\rangle\langle a_n|\psi_S\rangle\langle \psi_S|a_{n'}\rangle  \nonumber \\ &&=\sum_{n,n'}\delta_{n,n'}e^{-\lambda_A\frac{(a_n-a_{n'})^2}{2}} \langle a_n|\psi_S\rangle\langle \psi_S|a_{n'}\rangle  \nonumber \\ && = \sum_{n}e^{-\lambda_A\frac{(a_n-a_{n})^2}{2}} \langle \psi_S|a_{n}\rangle \langle a_n|\psi_S\rangle  \nonumber \\ && = \sum_n\langle \psi_S|a_{n}\rangle \langle a_n|\psi_S\rangle \nonumber \\ && =\langle \psi_S|\psi_S\rangle=1
\end{eqnarray}

Let us calculate now mean value of the measurements obtained from detector A  (note that first we detect A and afterwards immediately B):

 \begin{eqnarray}
&&\int_{-\infty}^{+\infty} a P_{\lambda_A,\lambda_B}^{\Sigma_B}(a) da=\\ && \int  a P_{\lambda_A,\lambda_B}(a,b)da db= \nonumber \\ &&(\frac{2\lambda_A}{\pi})^{1/2}\sum_{n,n'}\langle \psi_S|a_{n'}\rangle \langle a_n|\psi_S\rangle\times \nonumber \\ && [ \int da e^{-\lambda_A[(a-a_n)^2+(a-a_{n'})^2]}]\langle a_{n'}|a_n\rangle = \nonumber \\  &&(\frac{2\lambda_A}{\pi})^{1/2}\sum_{n,n'}\langle \psi_S|a_{n'}\rangle \langle a_n|\psi_S\rangle \times \nonumber \\ && [ \int da e^{-\lambda_A[(a-a_n)^2+(a-a_{n'})^2]}]\delta_{n,n'} =\nonumber \\ &&(\frac{2\lambda_A}{\pi})^{1/2}\sum_{n} [\int_{-\infty}^{+\infty} a e^{-2\lambda_A(a-a_n)^2}] |\langle a_n|\psi_S\rangle|^2 =\nonumber \\ &&\sum_{n} a_n |\langle a_n|\psi_S\rangle|^2= \langle \psi_S|\hat A|\psi_S\rangle \nonumber
\end{eqnarray}

 \textit{Here we prove that regardless of the values of the measurement strength parameters the mean measured value of A is the expectation value as calculated by quantum mechanics without taking into consideration the interaction of the system under study with the detectors. Another point that should be emphasized here is that this result is invariant under the uncertainty relations of $\hat A$ and $\hat B$.
}\\
Let us now calculate the mean value of B after taking the detections of A and B (first detect A and afterwards immediately B),
 \begin{eqnarray}
 \label{AbeforeB}
&&\int_{-\infty}^{+\infty} b P_{\lambda_A,\lambda_B}^{\Sigma_A}(b) db=\\ && \int  b P_{\lambda_A,\lambda_B}(a,b)da db= \nonumber \\ &&(\frac{2\lambda_B}{\pi})^{1/2}\sum_{n,n'}\langle \psi_S|a_{n'}\rangle \langle a_n|\psi_S\rangle e^{-\lambda_A\frac{(a_n-a_{n'})^2}{2}} \times \nonumber \\ && \sum_{m}[\int_{-\infty}^{+\infty}b e^{-\lambda_B(b-b_m)^2} db]\langle  b_m|a_n\rangle \langle a_{n'}|b_{m}\rangle = \nonumber \\ &&  \sum_{m} b_m \sum_{n,n'}e^{-\lambda_A\frac{(a_n-a_{n'})^2}{2}}\langle  b_m|a_n\rangle  \langle a_n|\psi_S\rangle\langle \psi_S|a_{n'}\rangle\langle a_{n'}|b_m\rangle =\nonumber \\&&\sum_{n,n'}e^{-\lambda_A\frac{(a_n-a_{n'})^2}{2}}\langle\psi_S|a_{n'}\rangle\langle a_{n'}| \sum_{m}b_m | b_m\rangle\langle  b_m|a_n\rangle  \langle a_n|\psi_S\rangle =\nonumber \\ && \sum_{n,n'}e^{-\lambda_A\frac{(a_n-a_{n'})^2}{2}}\langle\psi_S|a_{n'}\rangle\langle a_{n'}|\hat B |a_n\rangle  \langle a_n|\psi_S\rangle= \nonumber \\ && \sum_n \langle\psi_S|e^{-\lambda_A\frac{(a_n-\hat A)^2}{2}}\hat B |a_n\rangle  \langle a_n|\psi_S\rangle =\nonumber \\ &&\sum_n \langle\psi_S|\hat B e^{-\lambda_A\frac{(a_n-\hat A)^2}{2}}+ \nonumber \\ && [ e^{-\lambda_A\frac{(a_n-\hat A)^2}{2}},\hat B]|a_n\rangle\langle a_n |\psi_S\rangle =\nonumber \\ && \langle \psi_S|\hat B|\psi_S\rangle + \sum_n\langle \psi_S|[ e^{-\lambda_A\frac{(a_n-\hat A)^2}{2}},\hat B]|a_n\rangle\langle a_n |\psi_S\rangle \nonumber
\end{eqnarray}

Upon strong measurement, where $\lambda_A\to \infty$, the two measurements are still correlated

\begin{eqnarray}
&&\int_{-\infty}^{+\infty} b P_{\lambda_A\to\infty,\lambda_B}^{\Sigma_A}(b) db=\sum_n |\langle a_n|\psi_S\rangle|^2\langle a_n|\hat B|a_n\rangle \nonumber \\
\end{eqnarray}

  In the limit to weak measurements of A where $\lambda_A\to 0 $  (before taking the measurement of B)  the quantum correlations  between the two measurements is proportional to  $\lambda_A$

 \begin{eqnarray}
 \label{Effect-on-B}
&&\int_{-\infty}^{+\infty} b P_{\lambda_A\to 0,\lambda_B}^{\Sigma_A}(b) db - \langle \psi_S|\hat B| \psi_S\rangle \to  \\ &&   \frac{\lambda_A}{2}\sum_n\langle \psi_S|[\hat B,\hat A^2] + 2a_n[\hat A,\hat B]|a_n\rangle\langle a_n |\psi_S\rangle +... \nonumber
\end{eqnarray}

A simple case  is 1D problem where $\hat A=\hat x$ and $\hat B= \hat p_x$. In this case the quantum correlation effects are washed out in spite of the fact that $[\hat x,\hat p_x]\ne 0.$  The proof for this claim is straightforward when box quantization
is used. The discrete  spectrum of the position operator is $a_n\equiv x_n=\Delta_x\times n$ where $n=0,\pm 1,\pm 2,...$  and it is associated with a set of orthonormal functions $|a_n\rangle =sinc((x-x_n)/\Delta_x)$, where $sinc(\xi)=[\sin(\pi \xi)]/(\pi \xi)$. In such a case $\hat x| a_n\rangle =x_n|x_n\rangle$.  The commutation relations that determines the quantum interferences on the mean value of B (momentum in this example) as weak measurement of A is taken are given by,
\begin{eqnarray}
&&[e^{-\frac{\lambda_A}{2}(\hat x-x_n)^2},\hat p_x]=i\hbar \frac{\partial}{\partial x}e^{-\frac{\lambda_A}{2}(\hat x-x_n)^2}=\nonumber \\ && i\hbar \lambda_A|x-x_n|e^{-\frac{\lambda_A}{2}(x-x_n)^2}
\end{eqnarray}
Consequently,
\begin{eqnarray}
&&[e^{-\frac{\lambda_A}{2}(\hat x-x_n)^2},\hat p_x]|x_n\rangle= \\ && i\hbar \lambda_A|x-x_n|e^{- \frac{\lambda_A}{2}(x-x_n)^2}|x_n\rangle=0\nonumber
\end{eqnarray}
and the effect of the quantum correlations on the measurements of a mean value of $p_x$ (B) because of the weak measurement of A ($x_n$)  vanishes in spite of the fact that $[\hat x,\hat p_x]\ne0$. However, the effect of the quantum correlations will not be washed out if instead of measuring the momentum one will measure the distribution of the momentum since $[e^{-\frac{\lambda_A}{2}(\hat x-x_n)^2},\hat p^2_x]|x_n\rangle\ne 0$. Similar option in its nature is to detect in the second measurement not the momentum but the energy of the system where the operator that represents the second measurement is the Hamiltonian,  $\hat B=\hat H$
Another possibility is to keep the measurement of the momentum but instead of measurement of the position in the entire space is to measure it in a limited region as for example when replace the measurement of $\hat x$ by the measurement of $\hat V $, where the potential V in matrix representation is given by $V_{n,n'}=\langle x_n|\hat V|x_{n'}|\rangle$ and $a_n=V(x_n)$. Another possibility is  to have a narrow window on the measurement of the position as happen when the Husimi\cite{Husimi} distribution of a function is taken to measure the probability to allocate a quantum particle in a given point in the classical phase space where the measurement on the position is focused on the tail of the wavepacket that describes the quantum system under study(see Ref.\cite{Berry-NM}). As pointed out in Ref.\cite{NHQM-book} the tail of a time dependent functions is dominated by out going waves which fits well with case A in Ref.\cite{Berry-NM}. \\

Perhaps the results obtained in experiments carried our recently in the group of Yaron Zilberberg in Weizmann can be used to  support our findings\cite{YS}. Two photons are transferred trough  a lattice that is made of identical optical waveguides (WGs),
each supporting a single transverse mode. Two measurements were carried out. The first measurement (A in our discussion given above) is of the position of the in coming two photons at the entrance to the WGs (denoted as$A(z=0)$ where $z$ is the light propagation axis). In the second measurement (B in our notation)   the position of the outgoing photons is detected at a point along the propagation axis which is denoted as $z=z_f$. Therefore $B\equiv A(z_f)$. Since the index of refraction is not varied along the propagation axis the paraxial approximation holds and the Maxwell scalar equation that describes the light propagation in a lattice of identical array of optical WGs is similar to the time dependent Schro\"odinger equation where $z=time$. Therefore, even if it is the same operator $\hat A(z=0)$ and $\hat A(z_f)$ (different "times") do not commute. See a similar situation in the solution of the time-dependent Schr\"odinger equation in Ref.\cite{Belzig3}. The strong measurement of A implies that the uncertainty in the measured position of the two photons is small. As small as possible in these type of experiments. Namely, the two photons pass through the same single "slid" (optical waveguide). Weak measurement implies larger uncertainty in the measured position of the incoming two photons. It implies that the two photons can pass through two optical waveguides (WGs) that are far apart without our  ability to know through each one of the two WGs they have passed. The inverse of the distance between the two WGs is the strength measurement parameter ($\lambda_A$). When the two photons pass the same WG then $\lambda_A=\infty$. Quantum correlation maps obtained by the second measurement of the photons at the exit from the WGs were found to be very sensitive to the distance between the two WGs. The most dramatic effect was obtained when the two photons were injected into two adjacent WGs. That is, when $\lambda_A=1/D$ where $D$ is the distance between two adjacent WGs. \\

Another  simple possible experiment to test our theory is a transition of light (electrons, atoms)  through a slit that its width can be controlled.
The two-state pointer is "transmitted" versus "reflected".
By varying its width (perhaps not with sharp edges but with a gaussian transmission profile) we determine that accuracy of the measurements of the positions of the transmitted photons (particles).  The second type of measurements are of the wavelengths (momentum or spins)  of the photons (particles) after/before they passed through the slit. As before the degree of the uncertainty in the results of these measurements determine the strength of the measurements. The measurement of the positions of the photons/particles are associated with the measurements of detector A in our studies  given above, while the measurements of the wavelengths/momentum are associated with the measurements of detector B. Now we can do sudden sequential measurements and test our theoretical results presented above for different type of possibilities. Based on the derivation presented here one should measure positions and wavelengths/momenta/spins and show that although strong fluctuations in the measurements can be obtained upon weak measurements the mean values of these quantities are as expected from the calculations of the expectation values (assuming that the initial wavepacket distribution in position and wavelength/momentum/spin is known). Then show that the mean values of the square of positions or momentums/spins (wavelengths) are effected by the order of measurements (whether the positions were measured before or after the measurements of the wavelengths/momenta were taken). Also study the effect of the strength of the measurements as are defined by the corresponding
uncertainties on the obtained results. We can consider this set up as a two slit experiment (rather than the well known double slit experiment) where one slit is in spatial space and the other slit in momentum space.\\
 The experiment proposed above can be done in ultracold quantum gases in optical lattices. Spin-mixtures are loaded into the lattice potential. See the measurements taken in the Labs of Lukin\cite{Lukin} and Bloch\cite{Bloch1}-\cite{Bloch3}. The first weak measurement is the number of atoms which are trapped in the optical traps. Using a camera with a wavelength which is smaller than the lattice spacing each one of the measurements determine the number of atoms in one site in the optical lattice as described in  Ref.\cite{Bloch3}. If they will increase the wavelength of the camera then they will measure the number of atoms that occupy several sites. By repeating on single shot measurements one can get the average occupation of the atoms per one site. The ratio between the wavelength of the camera and the lattice spacing is the measurement strength parameter $\lambda_A$ is our derivation. As this ration is larger the strength of the measurement is weaker. On the basis of our proof given above the average number of atoms per site will be $\lambda_A$ independent (i.e., unaffected by the wavelength of the camera) while the standard deviation (fluctuation) of the measurements will be increased as $\lambda_A$ is increased. Last but not least is After taking measurements of the number of atoms per site you  another measurements can be taken. As for example measurement of the kinetic energy distribution of atoms as the standing laser waves are shut down suddenly and let the atoms move freely in space. This is a standard TOF (time of flight imaging), that can be carried out even if the  the setup in the experiments described in Ref.\cite{Bloch3} system does not allow to combine with the high resolution measurement of the kinetic energy release which is defined here as the second experiment. The high resolution of the measurement of the kinetic energy distribution is associated with either strong or weak  measurements. Since we have proved that  measured average value of the kinetic energy of the atoms in the optical lattice does not depend on the strength of the measurement and since the quality of the resolution is associated with the strength of the second measurement it seems that the experimental setup as described above is feasible. Moreover one can use this setup when the first measurement is measuring the number of atoms in one site at a given time $T_1$ after the initial preparing of one atom only in every one of the sites in the optical lattice.  The number of the atoms in one site in the optical lattice are measured as function of the wavelength of the camera. Then after some times at $t=T_2$ measuring again the number of atoms in one site when the wavelength of the camera is smaller than the optical lattice spacing. Because of the dynamical tunneling the measurements at time $t=T_2$ and time $t=T_2$ do not commute and therefore the conclusions based on our proof given above hold.

-----------------------

 The wavefunction which describes the distribution of the atoms in the different local minima in the 3D optical traps, $|(x_1,y_1,z_1),...(x_i,y_i,z_i)...\rangle$, is a linear combination of all possible distributions of the atoms in the different local minima (site) in the optical trap. The  square of the absolute values of the coefficients in this expansion provide the probability to have N number of atoms in a specific local minima in the 3D optical trap with a random distribution of spin. By using a camera with light with wavelength which is smaller than the distance between two adjacent local minima/site in the optical trap one can count the number of atoms which are located in one of the local minima/site in the trap. By repeating over this measurements many times one can calculate the probability to trap N numbers of atoms in one of the local minima (site) in the 3D optical trap. This is a strong measurement and the distribution of number of atoms in a specific local minima in the optical trap is equal to the distribution calculated from the linear coefficients in the expansion of the quantum wavepackets as mentioned above. The measured mean value of the number of atoms in one of the sites  in the 3D optical lattice is equal to the value obtained from the calculation of the expectation value of number of atoms using $|(x_1,y_1,z_1),...(x_i,y_i,z_i)...\rangle$.  Weak measurement are taken when the camera is a light beam that its wavelength is larger than the distance between two adjacent local minima in the 3D optical trap. As the wavelength of this light beam is larger the measurement of number of atoms in this measurement is weaker.  The ratio between the wavelength of the light beam of the camera in the first measurement and  the distance between two adjacent local minima in the trap can be considered as the strength measurement parameter $\lambda_A$ as appear in our derivations given above. \emph{Following our proof given in this paper the  mean value of the number of atoms in one local site   as obtained from the data obtained from the weak or the strong measurements will be as obtained from the calculated expectation value. The strength of the measurement (weak or strong) will not effect on the mean value of number of atoms in a given local minima in the 3D optical trap.} The second measurement we propose to do right after the measurement of the number of atoms in a given local minima in the tarp is  a strong measurement of the spin distribution of the atoms which are located in one of the local minima in the 3D optical trap. \emph{The number of the atoms  and the spin operators do not commute (while the position and spin of the particles commute) since the number of atoms in the site determines the energy of the atoms in this site
and Hamiltonian and spin do not commute because of spin-orbit coupling. Therefore based on our proof given above, the measurement of the mean value of the spin of the atoms and the standard deviation of the measured spin in one of the local minima in the 3D optical trap will be strongly effected by the variation of the wavelength of the camera's light beam in the first measurement of the number of atoms in a given local minima in the optical trap}.
 \\

We can conclude that here we have shown that by carrying out two successive measurements the mean  value of the measured results obtained for the second quantity can be very different from the expectation value as obtained in direct  strong measurements of this quantity. The conditions for it  are : (1) The two measurements are taken one after the other such the second measurement of B is taken when the system is in a state that was obtained resulting of the measurement of A.
(2) The two successive measurements of A and B are repeated many times when the system is initially prepared to be at the same $|\psi_S\rangle$ state. All detected values of $(a,b)$ are stored.
(3) The two measured quantities are presented by two non commuting operators, $[\hat A,\hat B]\ne 0$
(4) One should guarantee that   $[e^{-\frac{\lambda_A}{2}(\hat a -\hat A)^2},\hat B]|a_n\rangle\ne 0$ where $|a_n\rangle$ is an eigenstate of $\hat A$.
(5) The mean values of $a$ and $b$ which are calculated by using the experimental data are correspondingly compared with the expectation values $\langle \psi_S|\hat A|\psi_S\rangle$ and $\langle \psi_S|\hat B|\psi_S\rangle$.
(6) Based on our theoretical analysis it is expected that the mean value of $a$ will be exactly as obtained from the calculations of the expectation value $\langle a\rangle$, while the mean value of $b$ will deviate from $\langle b\rangle$. The deviation of the mean value of the second measurement will be effected by the strength of the measurement of B.
This is in harmony with Fuchs and Peres \cite{ASHER} claim that the act of observation of a state causes a distribution of this state. Here we have shown that by carrying out two succussive measurements one can control the broadness of this distribution. This might has a practical importance due to the
development of quantum cryptography.
\begin{acknowledgments}
 This research was supported in parts by the I-Core: the Israeli Excellence Center "Circle of Light", by the Israel Science Foundation grants No. 298/11 and No. 1530/15. I wish to thank Professor Abraham Nitzan for most simulating discussions on the use of Kraus operators to define the strength of a measurement, his enlighten comments on this work before submission, and  for the most pleasant and friendly hospitality during my two months visit in the University of Pennsylvania.
\end{acknowledgments}

 \end{document}